\documentclass[a4paper, 11pt]{article}

\usepackage[latin1]{inputenc}
\usepackage[T1]{fontenc}
\usepackage[english]{babel}
\usepackage{lmodern}
\usepackage{mathptmx}
\usepackage{times}
\fontfamily{times}

\usepackage{graphicx}
\usepackage{psfrag}
\usepackage{subfig}
\graphicspath{{./images/}}

\usepackage{geometry}
\usepackage{titlesec} 
\titleformat{\section}{\normalfont\small\bfseries}{\thesection.}{0.25em}{}
\titleformat{\subsection}{\normalfont\small\bfseries}{\thesubsection}{0.25em}{}

\usepackage{siunitx}
\usepackage[acronym]{glossaries}

\newacronym{lne}{LNE}{Laboratoire National de m\'etrologie et d\rq{}Essais}
\newacronym{l2s}{L2S}{Laboratoire des Signaux \& Syst\`emes}
\newacronym{cnrs}{CNRS}{Centre National de la Recherche Scientifique}

\newacronym{sur}{SUR}{Stepwise Uncertainty Reduction}
\newacronym{msur}{MSUR}{Maximum Speed of Uncertainty Reduction}

\newacronym{sl}{SL-SUR}{Single-Level SUR}

\usepackage{amsmath}
\usepackage{amssymb}
\usepackage{amsfonts}

\DeclareMathOperator*{\argmin}{argmin}
\DeclareMathOperator*{\argmax}{argmax}

\newcommand*{\ens}[1]{\mathbb{#1}}       
\newcommand*{\estim}[1]{\hat{#1}}        
\newcommand*{\refer}[1]{{#1}^{\star}}    
\newcommand*{\aver}[1]{\left<#1\right>}  

\newcommand*{\esp}{\mathbb{E}}              
\newcommand*{\var}{\mathbb{V}\mathrm{ar}}   
\newcommand*{\prob}{\mathbb{P}}             
\newcommand*{\measUnc}{H}          
\newcommand*{\p}{p}                         
\newcommand*{\globP}{P}                     
\newcommand*{\loss}{L}                      
\newcommand*{\Ltwo}{L^2(\mu)}               

\newcommand*{\N}{\mathsf{N}}                
\newcommand*{\GP}{\mathsf{GP}}              

\newcommand*{\zcrit}{z^{\mathrm{crit}}}     
\newcommand*{\Px}{\prob_{\ens{X}}}          
\newcommand*{\D}{\mathrm{d}}                
\newcommand*{\dt}{\Delta}
\newcommand*{\dx}{\D{x}}
\newcommand*{\dy}{\D{y}}
\newcommand*{\dmu}{\D{\mu}}
\newcommand*{\thf}{t^{\mathrm{HF}}}         
\newcommand*{\umax}{u_{\mathrm{max}}}       

\newcommand*{\widthFigure}{0.28\textwidth}
\newcommand*{\intFigure}{0.01\textwidth}

\usepackage{natbib}
\usepackage{apalike}

\newcommand{\templatefigures}
{\noindent
\begin{minipage}{2cm}
\begin{center}
  \centering
    \vspace{-1cm}
  \includegraphics[scale=0.1]{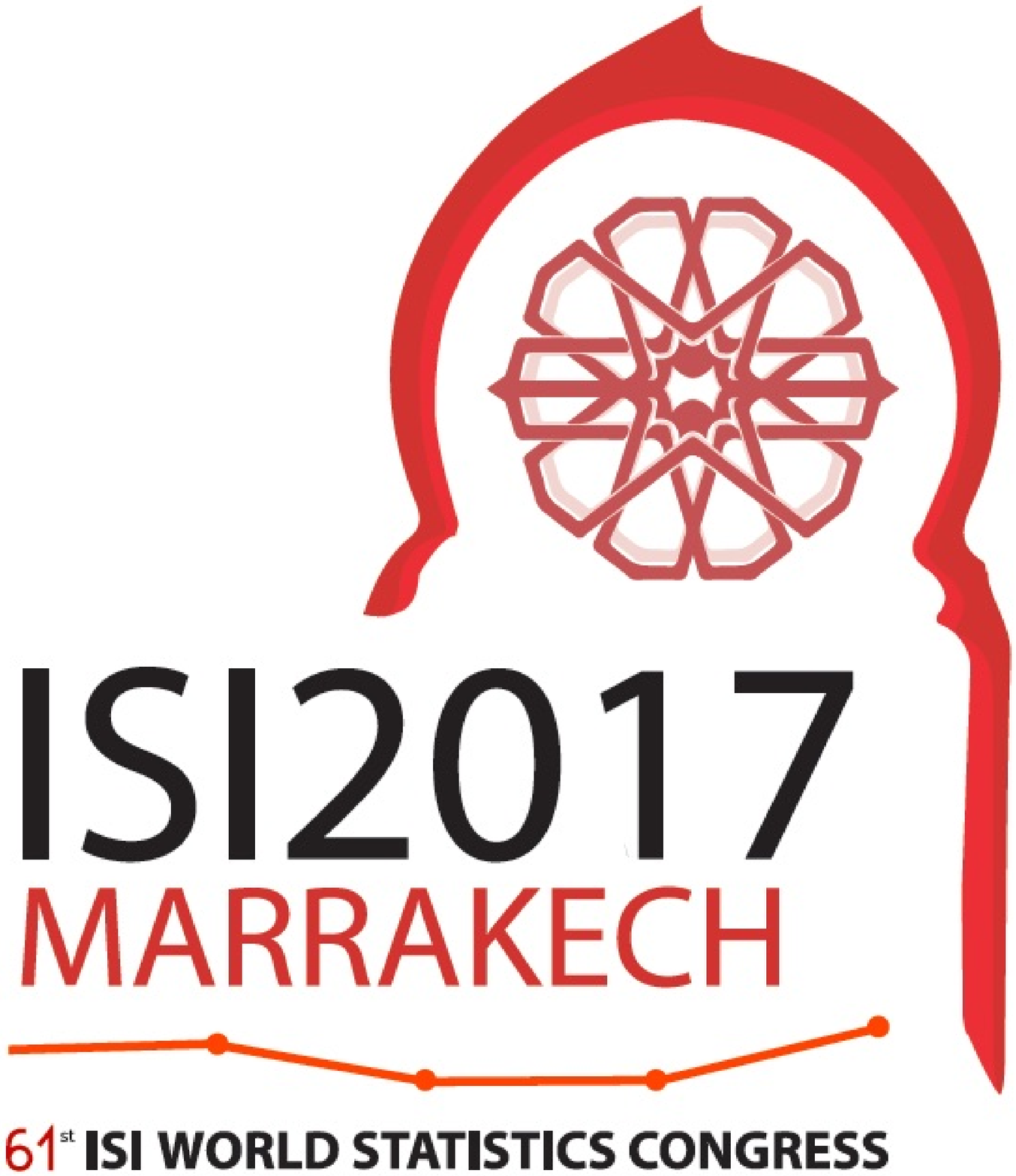}\\
\end{center}
\end{minipage}
\quad
\begin{minipage}{12cm}
\hspace*{6.8cm}
\end{minipage}
\quad
\begin{minipage}{2cm}
\begin{center}
\vspace{-0.9cm}
\includegraphics[scale=0.4]{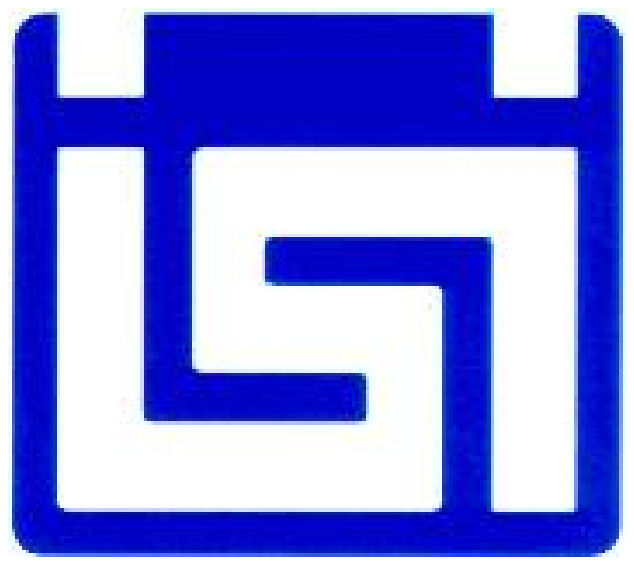}\\
\end{center}

\end{minipage}

\vskip0.2cm
}

\renewenvironment{abstract}
 {\small
  \begin{center}
  \bfseries \abstractname\vspace{-.5em}\vspace{0pt}
  \end{center}
  \list{}{%
    \setlength{\leftmargin}{0mm}%
    \setlength{\rightmargin}{\leftmargin}%
  }%
  \item\relax}
 {\endlist}

\begin{document}
\templatefigures{}

\begin{center}
  \textbf{Sequential design of experiments to estimate a probability
    of exceeding a threshold in a multi-fidelity stochastic simulator}
\end{center}

\begin{center}
  {R\'emi STROH*}\\
  {\acrfull{lne}, Trappes, France - remi.stroh@lne.fr}\\
  {\acrfull{l2s}, CentraleSup\'elec, Univ.~Paris-Sud,
    \acrfull{cnrs}, Universit\'e Paris-Saclay, Gif-sur-Yvette, France}\\
\vspace{0.5cm}

{S\'everine DEMEYER, Nicolas FISCHER}\\
{\acrshort{lne}, Trappes, France - name.surname@lne.fr}\\
\vspace{0.5cm}

{Julien BECT, Emmanuel VAZQUEZ}\\
{\acrshort{l2s}, CentraleSup\'elec, Univ.~Paris-Sud, \acrshort{cnrs},
  Universit\'e Paris-Saclay, Gif-sur-Yvette,
  France - name.surname@centralesupelec.fr}\\

\end{center}

\begin{abstract}
  In this article, we consider a stochastic numerical simulator to
  assess the impact of some factors on a phenomenon. The simulator is
  seen as a black box with inputs and outputs.  The quality of a
  simulation, hereafter referred to as fidelity, is assumed to be
  tunable by means of an additional input of the simulator (e.g., a mesh
  size parameter): high-fidelity simulations provide more accurate
  results, but are time-consuming. Using a limited computation-time
  budget, we want to estimate, for any value of the physical inputs, the
  probability that a certain scalar output of the simulator will exceed
  a given critical threshold at the highest fidelity level.
  The problem is addressed in a Bayesian framework, using a Gaussian
  process model of the multi-fidelity simulator. We consider a Bayesian
  estimator of the probability, together with an associated measure of
  uncertainty, and propose a new multi-fidelity sequential design
  strategy, called \acrfull{msur}, to select the value of physical
  inputs and the fidelity level of new simulations. The \acrshort{msur}
  strategy is tested on an example.\\

\end{abstract}

\section{Introduction}

The objective of this article is to propose a new Bayesian algorithm
for \emph{sequential design of experiments} in the context of
\emph{multi-fidelity stochastic} numerical simulators. %
A numerical simulator is a computer program modeling a physical
phenomenon or a system. When the simulator is deterministic, running
the computer program twice using the same inputs yields the same
outputs. %
In this case, the simulator can be viewed formally as a function. When
the simulator is stochastic, running twice the simulator with the same
inputs does not return the same output. %
Moreover, we assume that the simulator has a particular input called
\emph{fidelity parameter}---e.g., the mesh size of a finite-difference
partial differential equation solver---that controls a trade-off
between quality of simulation and computation time. %
A high-fidelity simulation provides an accurate result, but is
time-consuming.

Let~$t \in \ens{T}$ be the fidelity input of the simulator,
and~$x \in \ens{X}$ the vector of all other inputs. %
At fixed $(x,\, t)$, the output of the simulator $Z$ follows
a probability distribution $\prob^Z_{x,\, t}$. %
We denote by $\thf$ the value of the fidelity input associated to the
highest available level of fidelity: %
this is the level of interest for the user of the simulator. %
In this article, we focus on comparing the output $Z$ to a critical
threshold $\zcrit$---typically, a level of output that should not be
exceeded by the physical phenomenon or the system under study. %
This comparison can be studied from two points of view: either
locally, by considering the function~$\p$ that gives for each
point~$x$ the probability of exceeding the threshold at this point,
\begin{equation}
  \p\left(x\right) = \prob^Z_{x,\, \thf}\left(Z >
    \zcrit\right),\quad x\in\ens{X},
\end{equation}
or globally, by computing the global probability $\globP$ of exceeding
the threshold, which may be written as
\begin{equation} \label{eq:proba} %
  \globP = \int_{\ens{X}}\p\left(x\right) \Px(\dx),
\end{equation}
where $\Px$ is a probability distribution which models uncertainty
about the value of the input factors.

Our objective is to estimate these quantities by observing the
stochastic outcomes $z_1,\,z_2\dots$ of simulations at points
$(x_1,\, t_1),\, (x_2,\,t_2)\ldots$ %
The estimators are built using a Bayesian model of the simulator. %
More precisely, we assume a prior model about $\prob^Z_{x,\,t}$ and
compute the posterior distribution of the quantity of interest (either
$\p$ or~$\globP$) given
$\chi_n = \left(x_i,\, t_i;\, z_i\right)_{i\leq n}$. %
In this article, we suggest a new sequential design algorithm to
select points $(x_1,\, t_1),\, (x_2,\,t_2)\ldots$ in order to obtain a
\emph{fast} reduction of the uncertainty about~$\p$ or $\globP$. %
Our methods deals both with the stochastic nature of the simulator and
the tunable simulation cost.

The paper is organized as follows. %
Section~\ref{sec:goal} sets the Bayesian framework that we consider for
stochastic multi-fidelity simulators. %
Section~\ref{sec:algorithm} describes our sequential algorithm and our
new sampling criterion called \gls{msur}. %
Section~\ref{sec:illustrations} illustrates the method on a test
problem, and Section~\ref{sec:conclu} concludes the paper.

\section{Bayesian framework}
\label{sec:goal}

Following our previous work on multi-fidelity stochastic simulators
\citep{stroh2016gaussian}, assume that the output $Z$
of a simulation at~$(x,\, t)$ follows a normal distribution:
\begin{equation}
  Z \mid \xi, \lambda \,\sim\,
  \N\left(\xi (x,\, t), \lambda(x,\, t)\right)\,.
\end{equation}
Assume moreover that conditional on~$\xi$ and~$\lambda$, all runs of
the simulator are independent. %
Of course, the choice of a normal distribution here is a simplifying
hypothesis, that must be verified in practice for each particular
simulator%
\footnote{%
  Other (possibly non-parametric) families of distributions could be
  considered as well. %
  Note, however, that the convenient conjugacy property of the
  Gaussian process prior with respect to the Gaussian likelihood would
  be lost by doing so.}.

Assume a Gaussian process prior for the mean function~$\xi$:
\begin{equation}
  \xi \mid m, k
  \;\sim\; \GP\left(m,\,
    k\left((x,\, t), (x',\, t')\right)\right),
  \qquad m \;\sim\; \mathsf{U} \left(\ens{R}\right),
\end{equation}
where $m$ is the (constant) mean of the Gaussian process, $k$ its
covariance function, and $\mathsf{U}\left(\ens{R}\right)$ the
(improper) uniform distribution on $\ens{R}$. %
For the sake of simplicity, the functions $\lambda$ and $k$ are
assumed to be known. %
Under this prior, the posterior distribution of~$\xi$ conditionally
to the observations~$\chi_n$ is Gaussian:
\begin{equation}
  \xi \mid \chi_n \,\sim\, \GP\left(m_n(x,\, t),\,
    k_n\left((x,\, t), (x',\, t')\right)\right),
\end{equation}
where~$m_n$ and~$k_n$ are given by the kriging formulas \citep[see,
e.g.,][]{stein1999interpolation}.

Since $Z$ has a Gaussian distribution conditional on $\xi$ and
$\lambda$, for all $x\in\ens{X}$, the probability of exceeding a
critical value~$\zcrit$ can be written as
\begin{equation}
  \p\left(x\right) =
  \Phi\left(\frac{\xi\left(x,\, \thf\right) - \zcrit}
    {\sqrt{\lambda\left(x,\, \thf\right)}}\right),
\end{equation}
with $\Phi$ the cumulative distribution function of the normal
distribution. %
Then, denoting by $\esp_n = \esp\left(\,\, \cdot \mid \chi_n\right)$
and $\var_n = \var\left( \,\,\cdot \mid \chi_n\right)$ the posterior
mean and variance operators, we have: %
\begin{align}
  \esp_n\left(\p\left(x\right)\right)
  & = \Phi\left(u_n\left(x\right) \right),\\
  \var_n\left(\p\left(x\right)\right)
  & = \Phi_2\left(u_n\left(x\right), u_n\left(x\right);
    r_n\left(x, x\right)\right)
    - \Phi\left(u_n\left(x\right) \right)^2,
\end{align}
where $\Phi_2\left(\,\cdot, \cdot\,;\, r\right)$ stands for the
bivariate normal distribution function with correlation $r$,
$V_n(x,\, t) = \lambda(x,\, t) + k_n\left((x,\, t),
    (x,\, t)\right)$,
$u_n(x) = \frac{m_n\left(x,\, \thf\right) - \zcrit}
{\sqrt{V_n\left(x,\, \thf\right)}}$, and
$r_n(x,\, x') = \frac{k_n\left(\left(x,\, \thf\right),\left(x',\,
      \thf\right)\right)} {\sqrt{V_n\left(x,\, \thf\right) V_n\left(x',\,
      \thf\right) }}$.

\section{Proposed algorithm}
\label{sec:algorithm}

\subsection{A SUR criterion to estimate a probability of
  exceeding a threshold in the case of stochastic outputs}

The goal of sequential design of experiments is to select a simulation
point $\left(x_{n+1}, t_{n+1}\right)$, using the result of the previous
observations $\chi_n$, to obtain a good improvement of our knowledge
about a given quantity of interest.
Our algorithm is based on the idea of \gls{sur} strategies
\citep{vazquez2009sequential, bect2012sequential}. %
The main idea of SUR strategies is to define a measure of
uncertainty $\measUnc_n$ about the estimator of a quantity of
interest.
Then, assuming that there is no fidelity parameter~$t$,
the next observation point is selected in order to minimize the
expected uncertainty using this new observation, whose outcome is
random:
\begin{equation}
  x_{n+1} = \argmin_{x \in \ens{X}}\;
    \esp_n\left(\measUnc_{n+1}\mid
    X_{n+1} = x\right).
  \label{eq:surPrinc}
\end{equation}

Several measures of uncertainty $\measUnc_n$ have been proposed in the
literature \citep{bect2012sequential, chevalier2014fast,
  azzimonti2016adaptive} for the problem of estimating the volume or
contour of an excursion set of a deterministic function. %
In our framework, however, we deal with a stochastic simulator, and we
propose a new measure of uncertainty for this case.  Let $L$ be the
$L^{2}$ loss function
\begin{equation}
  \loss\left(f, g\right)
  = \left\|f - g\right\|^2_{\Ltwo}
  = \int_{\ens{X}} \left(f(x) - g(x)\right)^2 \mu(\dx),
\end{equation}
where $\mu$ denotes a positive measure on~$\ens{X}$. %
We suggest to measure uncertainty by the $L^{2}$
loss incurred when estimating $\p$ by $\estim{\p}_n = \esp_n(\p)$:
\begin{equation}
  \measUnc_n
  = \esp_n\left(
  \loss\left(\estim{\p}_n,~ \p\right)\right)
  = \int_{\ens{X}} \var_n\left( \p(x) \right) \mu(\dx).
\end{equation}
Note that in the case where the goal is to estimate~$\globP$,
$\measUnc_n$ provides an upper bound on the posterior variance for
any~$\Px$ that admits a density~$g$ with respect to~$\mu$: %
$\var_n\left(\globP\right) \le G\; \measUnc_n$, where
$G = \int g^2 \dmu$ and~$g = \D{\Px} / \dmu$.
Consequently, a sequential design strategy that aims to
reduce~$\measUnc_n$ will be useful not only to estimate~$\p$ itself
but also to estimate~$\globP$ for a large class of probability
distributions~$\Px$.

Let $J_n(x,\, t) = \esp_n\left(\measUnc_{n+1}\mid
  X_{n+1} = x, T_{n+1} = t\right)$. Since the output of the
simulator has a Gaussian distribution, $J_n\left(x,
  t\right)$ can be written as
\begin{equation}
  J_n(x,\, t) = \int_{\ens{X}}\Bigl(
  \Phi_2\bigl(u_n\left(y\right), u_n\left(y\right);
  r_n\left(y, y\right)\bigr)
  - \Phi_2\bigl(u_n\left(y\right), u_n\left(y\right);
  \frac{k_n\left(\left(y, \thf\right),
      (x,\, t)\right)^2}
  {V_n\left(y, \thf\right)V_n(x,\, t)}\bigr)
  \Bigr)\, \mu (\dy).
\label{eq:surStrat}
\end{equation}

\subsection{Dealing with tunable fidelity}

The \gls{sur} strategy~\eqref{eq:surPrinc} is relevant when there is
no fidelity parameter~$t$. %
In a multi-fidelity context, however, the simulator becomes more and
more time-consuming as the fidelity parameters comes closer
to~$\thf$. %
We propose a new sequential design algorithm which takes into account
the variable simulation cost.

Let $C(x,\, t) > 0$ be the cost of observing the simulator at
$(x,\, t)$.  The idea is to balance the benefit of an
observation against its cost.  For us, the benefit of an observation is
the reduction of uncertainty.  Hence, we propose the following
sequential strategy called \acrfull{msur}:
\begin{equation}
  \left(x_{n+1}; t_{n+1}\right) =
  \argmax_{(x, t) \in \ens{X} \times \ens{T}}\;
  \frac{\measUnc_n - J_n(x, t)}{C(x, t)}.
\label{eq:algo}
\end{equation}
This idea is similar to the one proposed by
\citet{huang2006sequential} in the context of optimization, where the
expected improvement is divided by the cost of an observation.
Likewise, for the purpose of providing a surrogate model of the simulator,
\citet{gratiet2015kriging} suggested to choose the new level
of fidelity by comparing the cost of the level and the
reward in terms of reduction of uncertainty.

Note that in many applications, the cost $C$ depends only on the level
$t$: $C(x,\, t) = C(t)$.  In this case,
\eqref{eq:algo} can be divided into two steps: first, select the
input $x$ at each level that minimizes the \gls{sur} criterion; second,
select the level $t$ that maximizes the \gls{msur} criterion.  Our
sequential design algorithm can thus be rewritten in this case as
\begin{equation}
   \left\{
     \begin{aligned}
       \tilde{x}(t) & =
       \argmin_{x \in \ens{X}}\; J_n(x, t),\\
       t_{n+1} & = \argmax_{t \in \ens{T}}\;
       \frac{\measUnc_n
         - J_n\left(\tilde{x}\left(t\right), t\right)}{C(t)},\\
       x_{n+1} & = \tilde{x}\left(t_{n+1}\right).\\
     \end{aligned}
   \right.
   \label{eq:MSUR}
\end{equation}

\section{Illustration}
\label{sec:illustrations}

We illustrate our algorithm on a two-dimensional example
inspired by \citet{au2001estimation}.  We consider a damped harmonic
oscillator subject to  random forcing, described by the stochastic
differential equation
\begin{equation}
  \ddot{X}\left(u\right) + 2\zeta\omega_0\dot{X}\left(u\right)
  + \omega_0^2X\left(u\right) = W\left(u\right),
  \quad X\left(0\right) = 0,
  \quad \dot{X}\left(0\right) = 0,\quad u \in \left[0; \umax= \SI{30}{\second}\right],
\label{eq:defX}
\end{equation}
where $W$ is a Gaussian white noise (with spectral density equal to
one).  We compute an approximation $(\estim{X}_n)_{n=0,\ldots,
  \lfloor\umax/\dt\rfloor}$ of $X$ by finite differences with a time
step $\dt$.  We use an explicit exponential Euler scheme \citep[see,
e.g.,][]{jentzen2009overcoming}. The output of our simulator is defined as
\begin{equation}
  \begin{array}{cccc}
    Z:
    & \left[0; 30\right] \si{\radian\per\second}
      \times \left[0; 1\right] \times \left]0; 1\right]\si{\second}
    & \to
    & \ens{R}\\
    & \left(\omega_0, \zeta, \dt\right)
    & \mapsto
    & \log\left(\max\left\{ \estim{X}_n, n
      \leq \left\lfloor\frac{\umax}{\dt}\right\rfloor\right\}\right).
  \end{array}
  \label{eq:function}
\end{equation}

\begin{figure}[t]
\centering
\psfrag{pulse (rad/s)}[tc][tc]{$\omega_0$}
\psfrag{damping}[bc][bc]{$\zeta$}
\psfrag{Mean function}{}
\psfrag{Standard deviation function}{}
\psfrag{Probability of exceeding the threshold -3}{}

\subfloat[$\xi\left(x, \dt = \SI{1}{\second}\right)$]
{\includegraphics[width = \widthFigure]{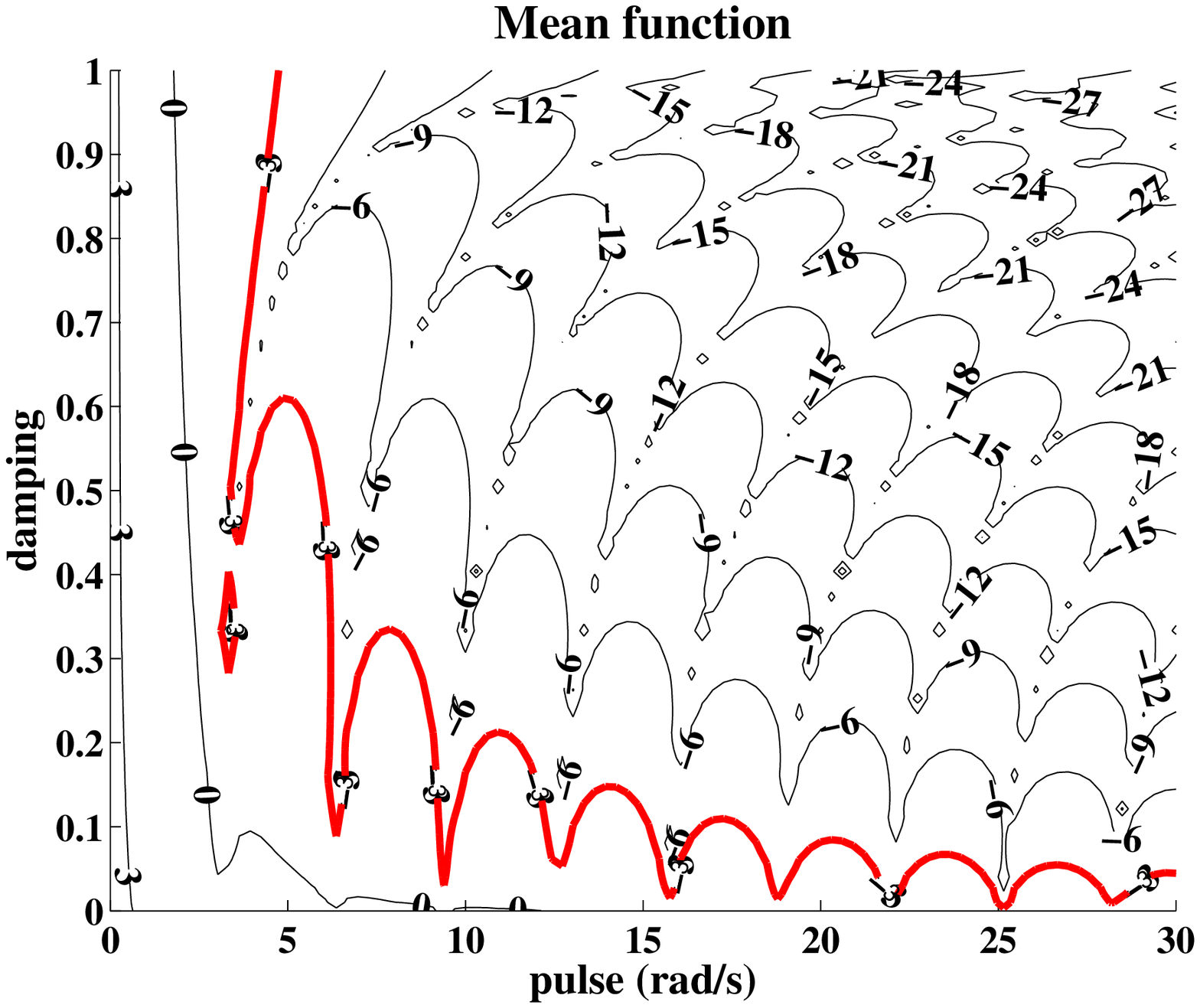}}
\hspace{\intFigure}
\subfloat[$\xi\left(x, \dt = \SI{0.1}{\second}\right)$]
{\includegraphics[width = \widthFigure]{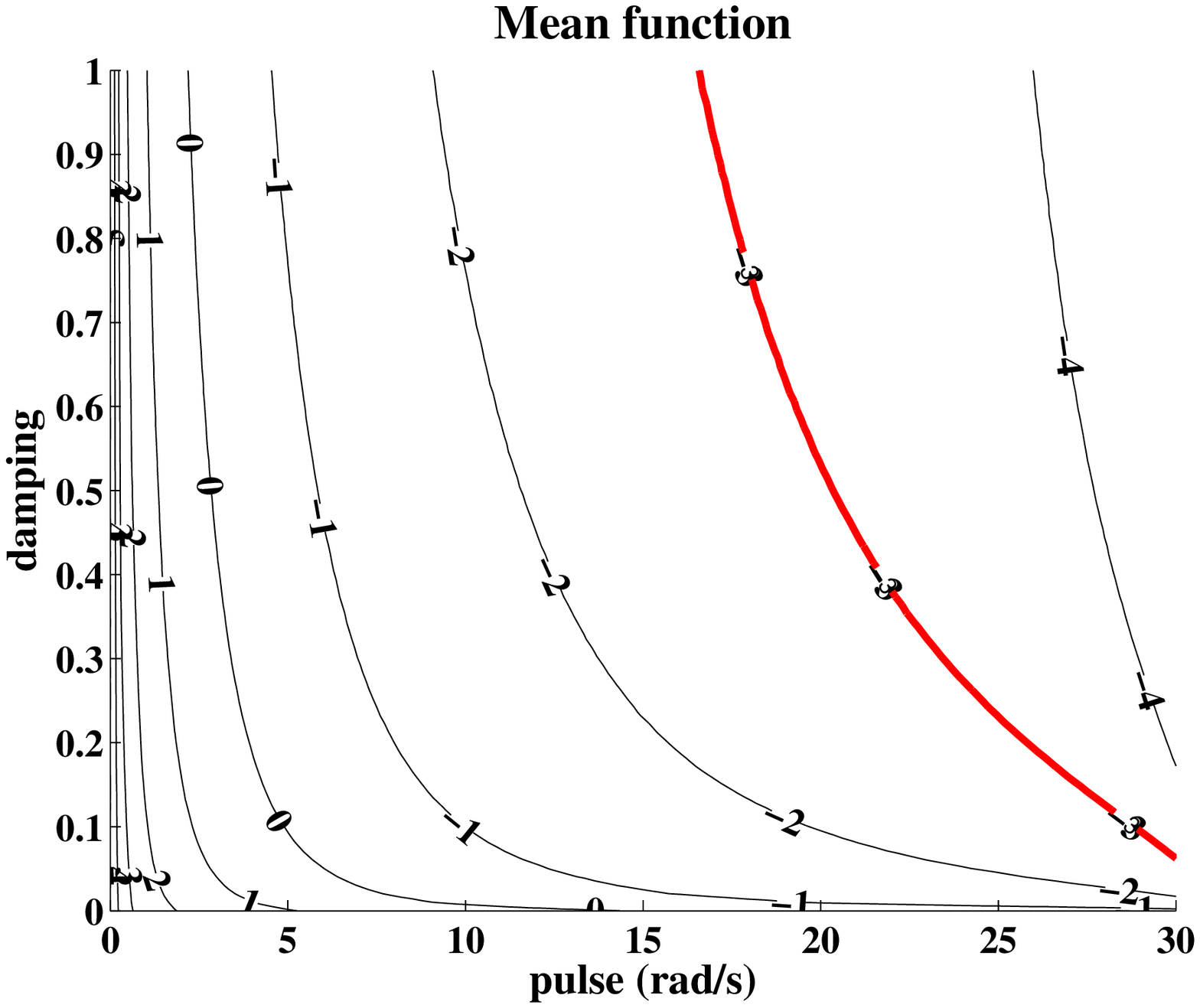}}
\hspace{\intFigure}
\subfloat[$\xi\left(x, \dt = \SI{0.01}{\second}\right)$]
{\includegraphics[width = \widthFigure]{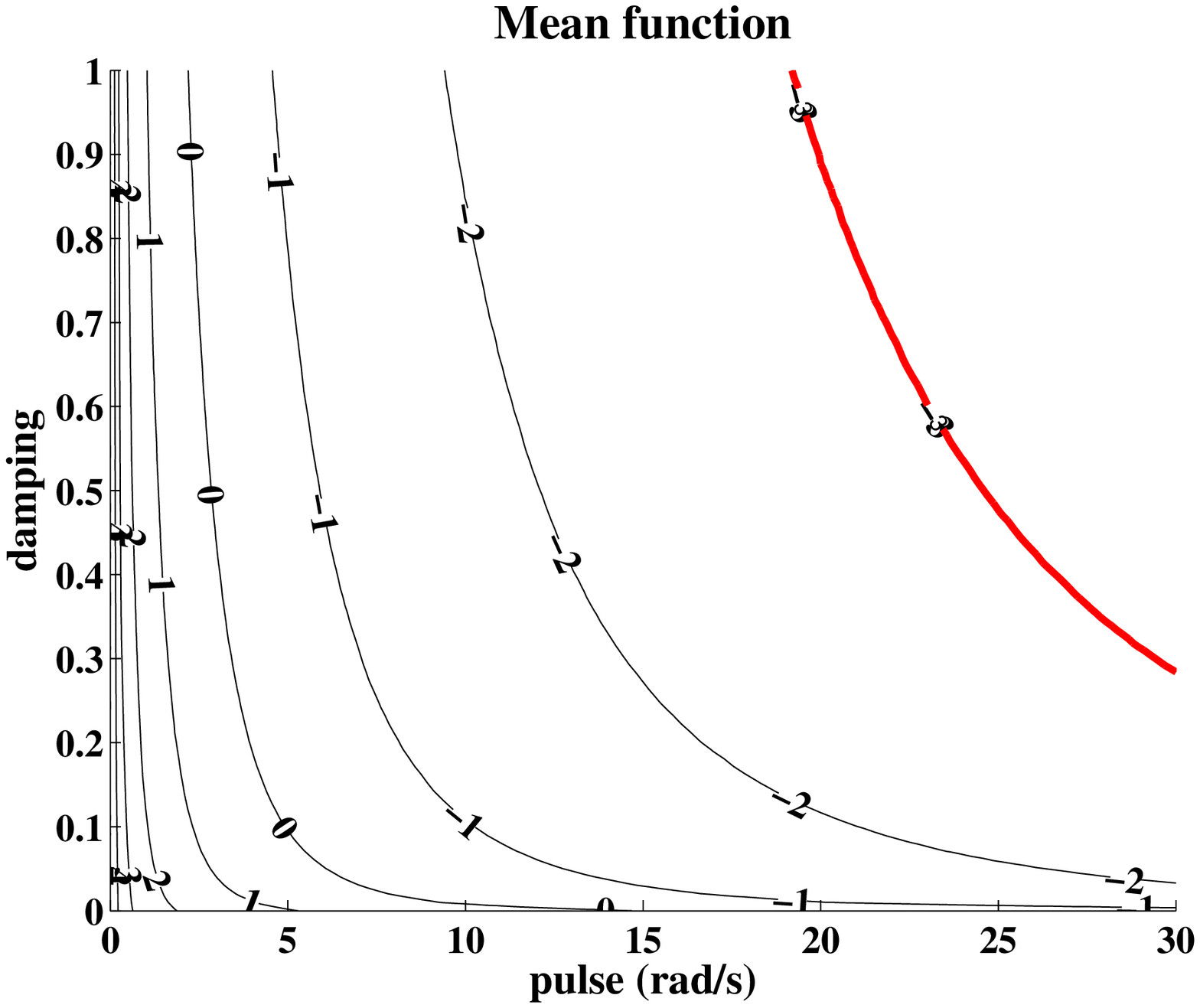}}

\subfloat[$\sqrt{\lambda\left(x, \dt = \SI{1}{\second}\right)}$]
{\includegraphics[width = \widthFigure]{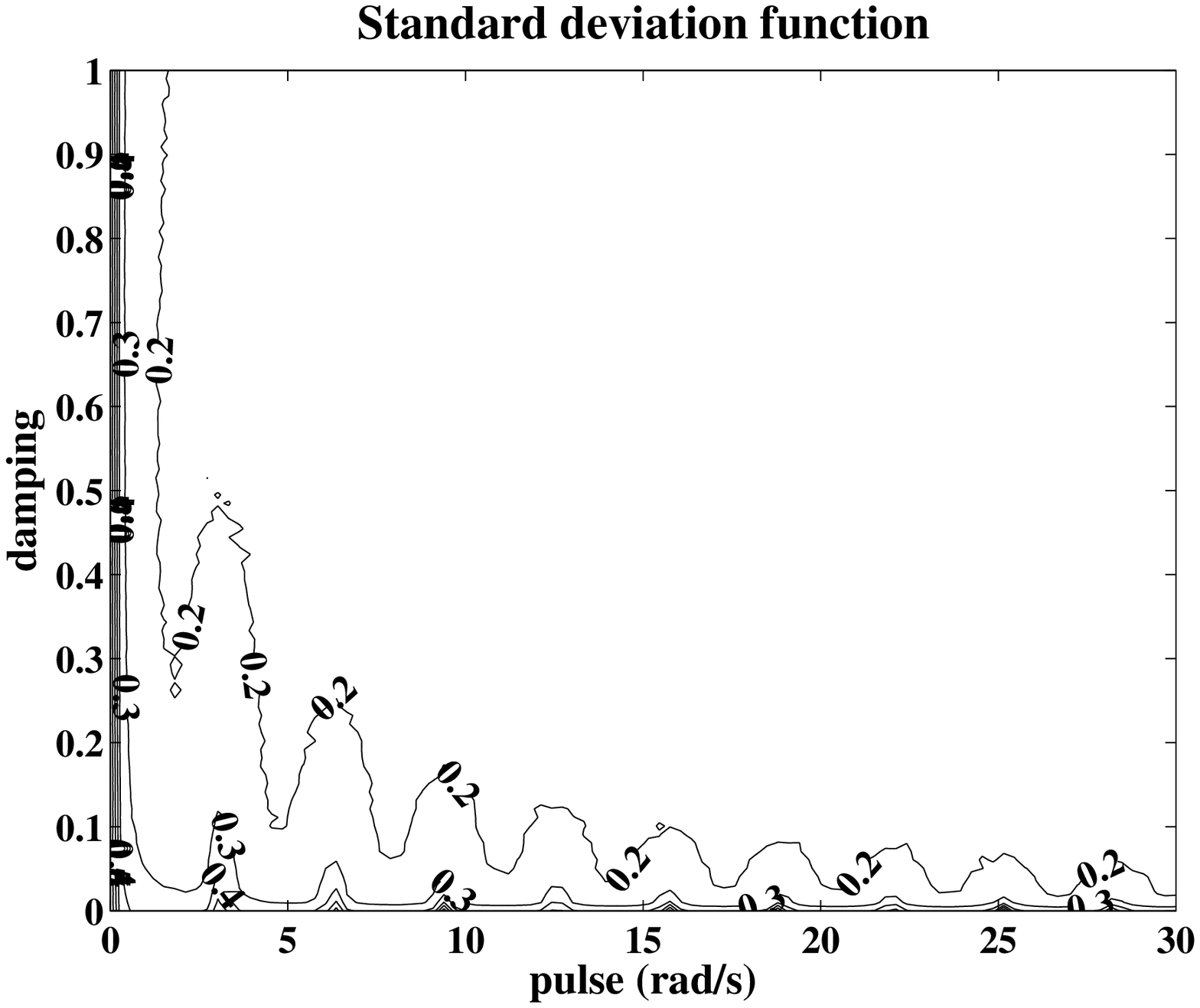}}
\hspace{\intFigure}
\subfloat[$\sqrt{\lambda\left(x, \dt = \SI{0.1}{\second}\right)}$]
{\includegraphics[width = \widthFigure]{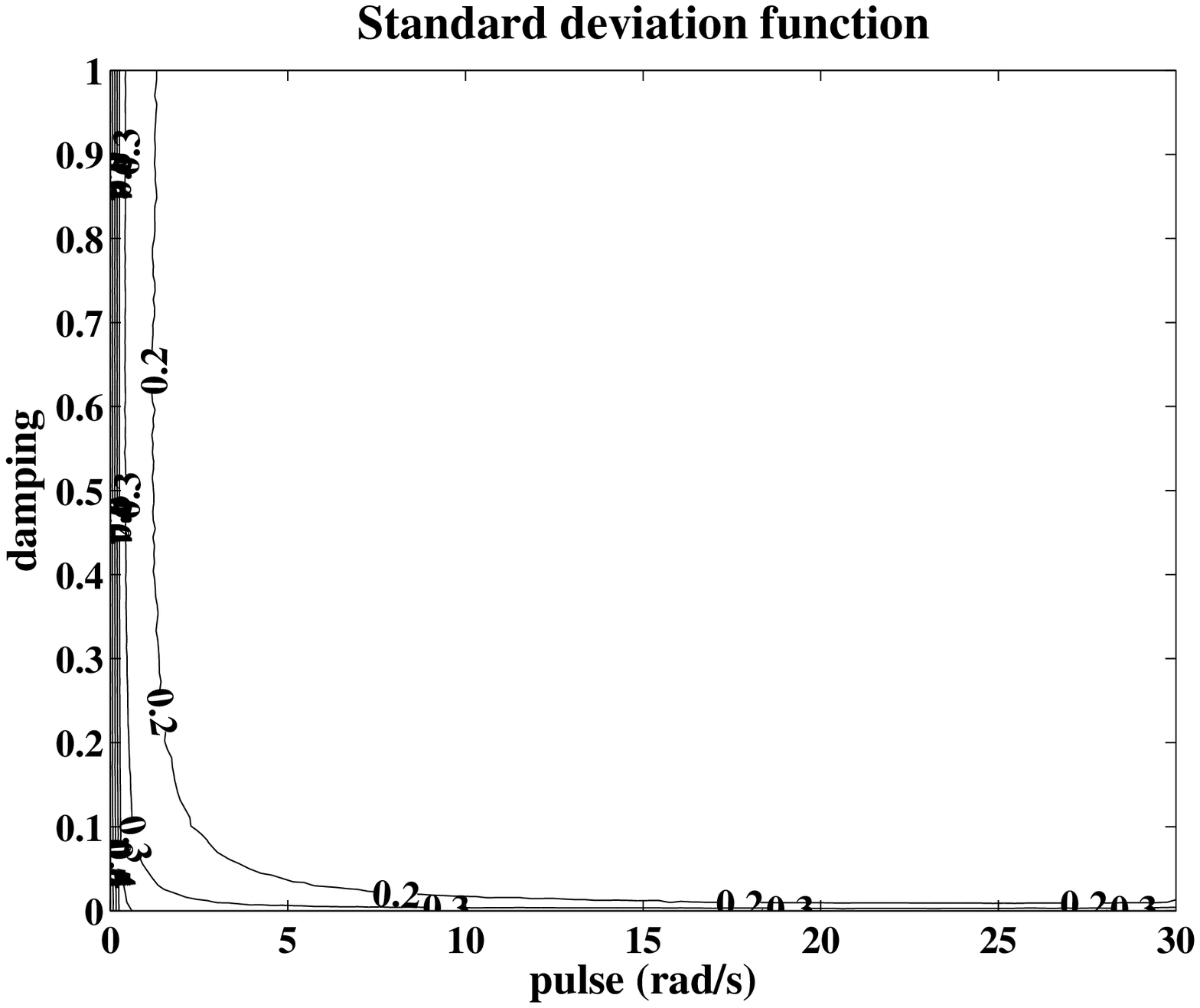}}
\hspace{\intFigure}
\subfloat[$\sqrt{\lambda\left(x, \dt = \SI{0.01}{\second}\right)}$]
{\includegraphics[width = \widthFigure]{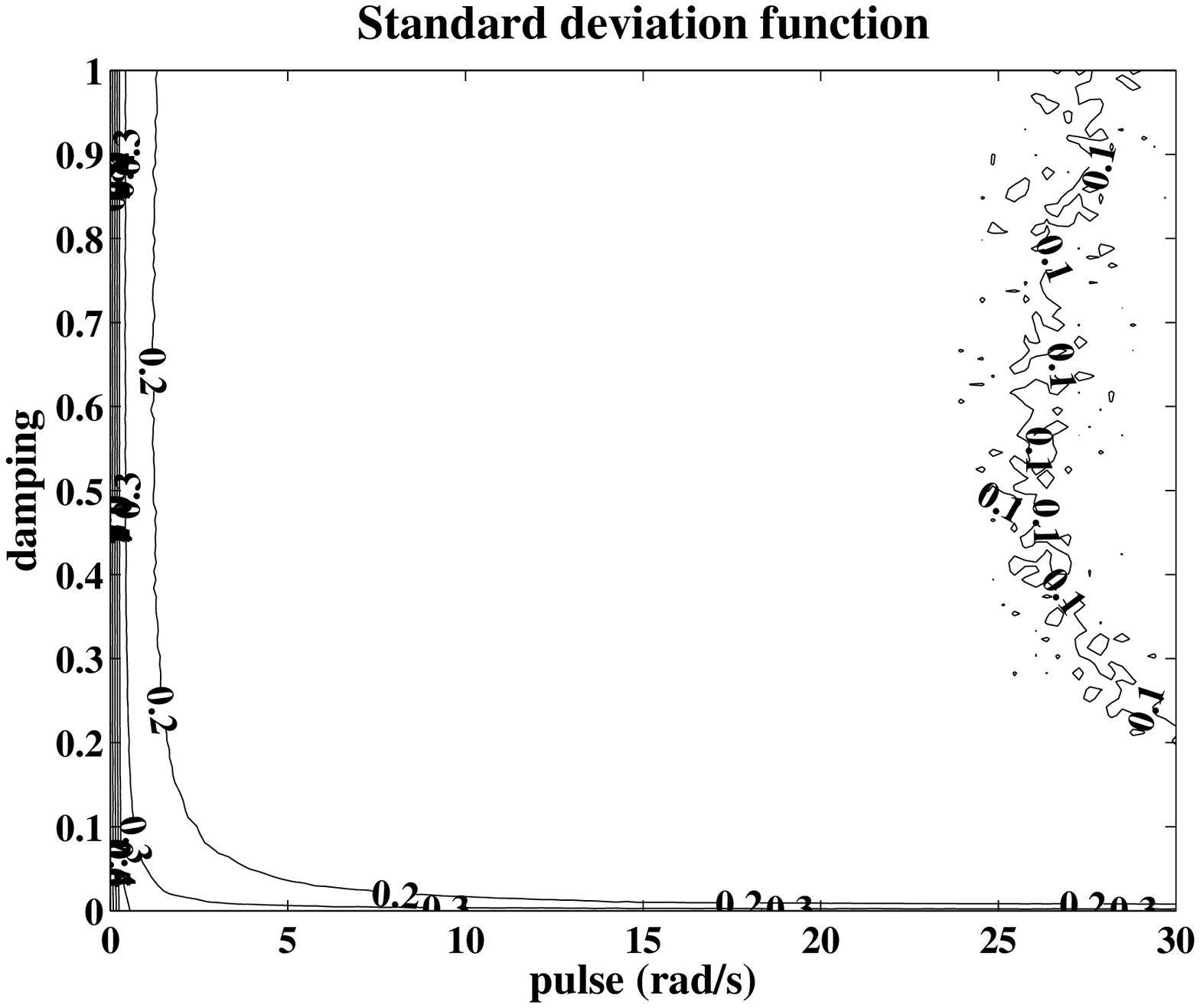}}

\subfloat[$\prob^Z_{x, 1}\left(Z > - 3\right)$]
{\includegraphics[width = \widthFigure]{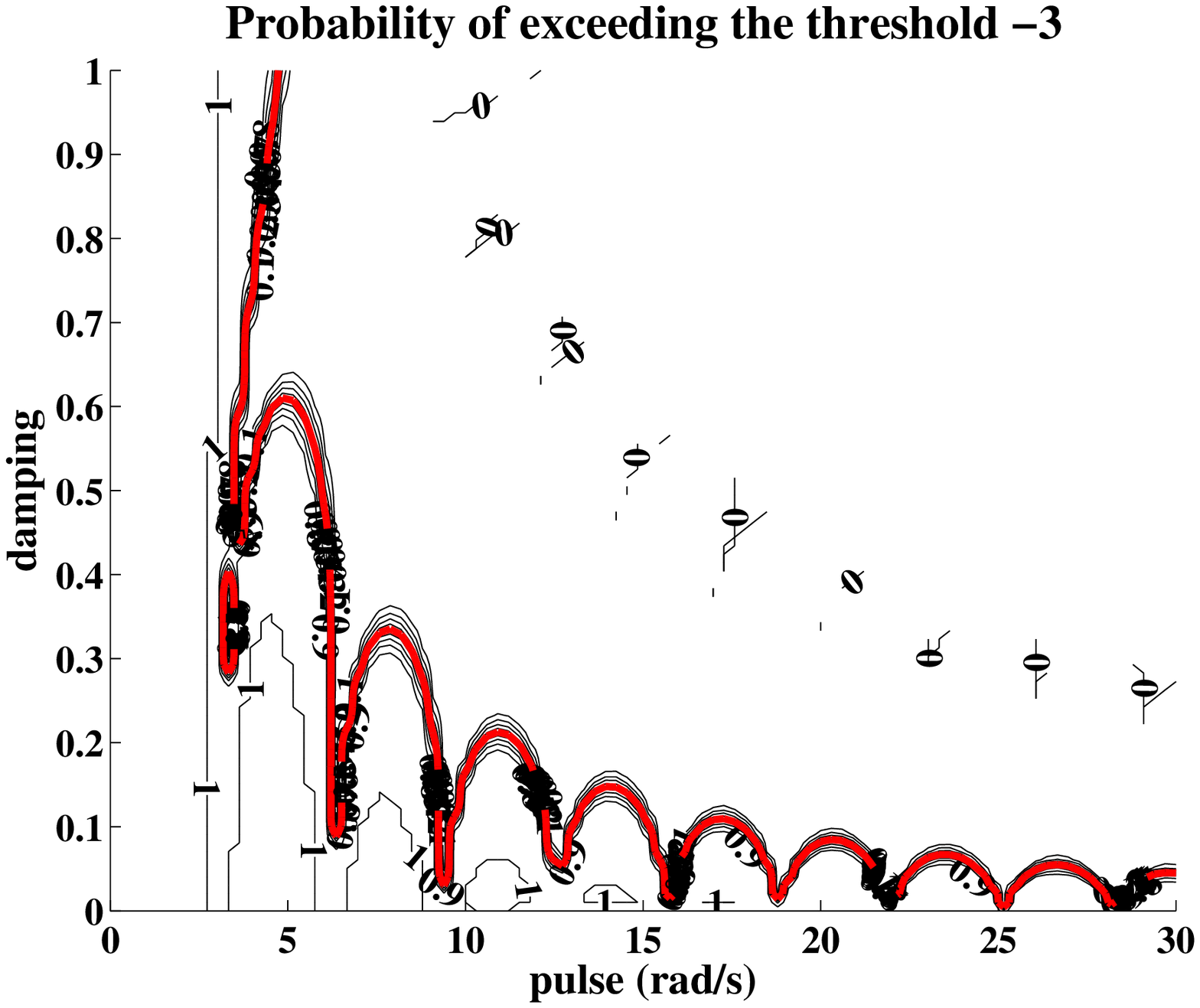}}
\hspace{\intFigure}
\subfloat[$\prob^Z_{x, 0.1}\left(Z > - 3\right)$]
{\includegraphics[width = \widthFigure]{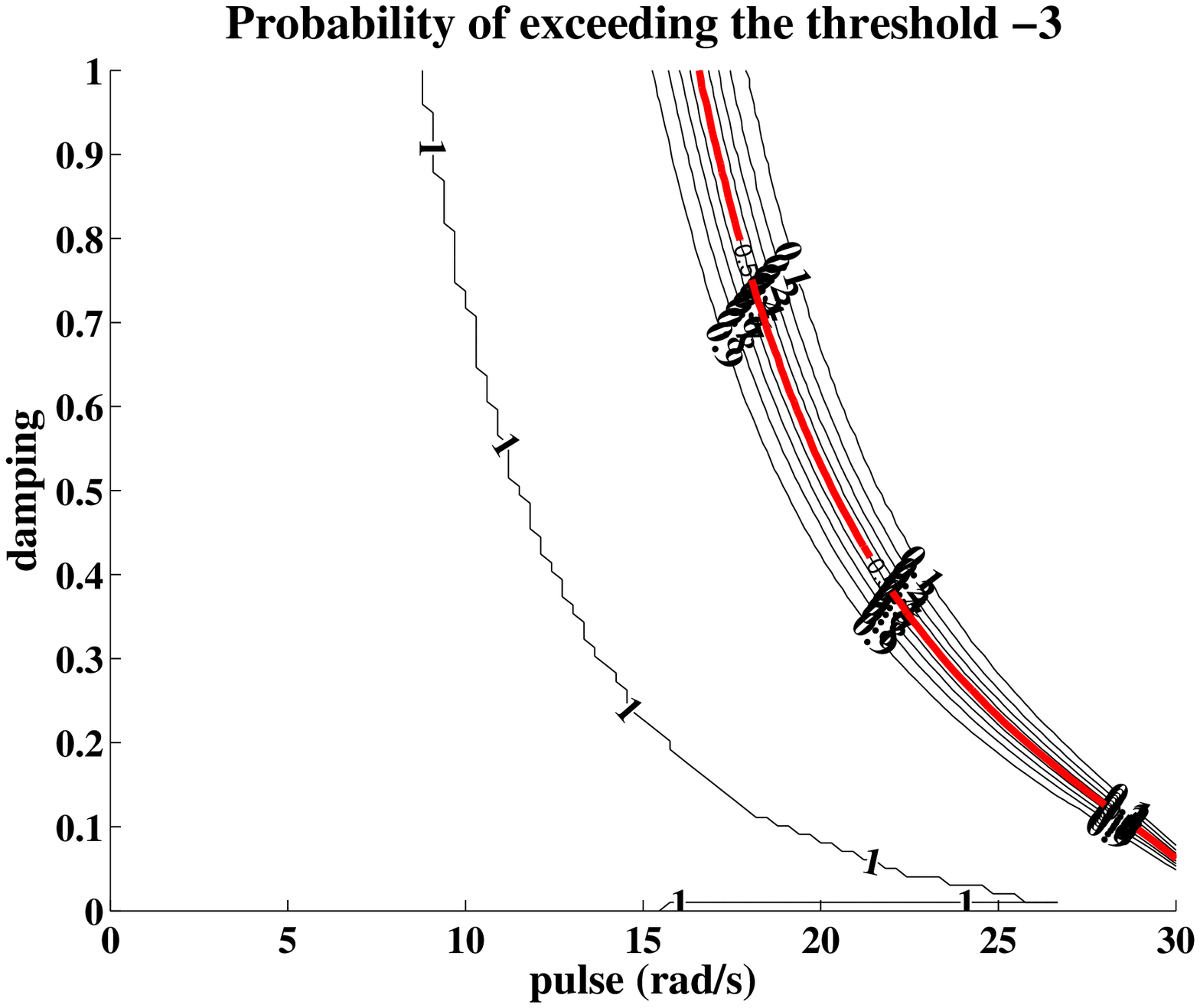}}
\hspace{\intFigure}
\subfloat[$\prob^Z_{x, 0.01}\left(Z > - 3\right) =\p\left(x\right)$]
{\includegraphics[width = \widthFigure]{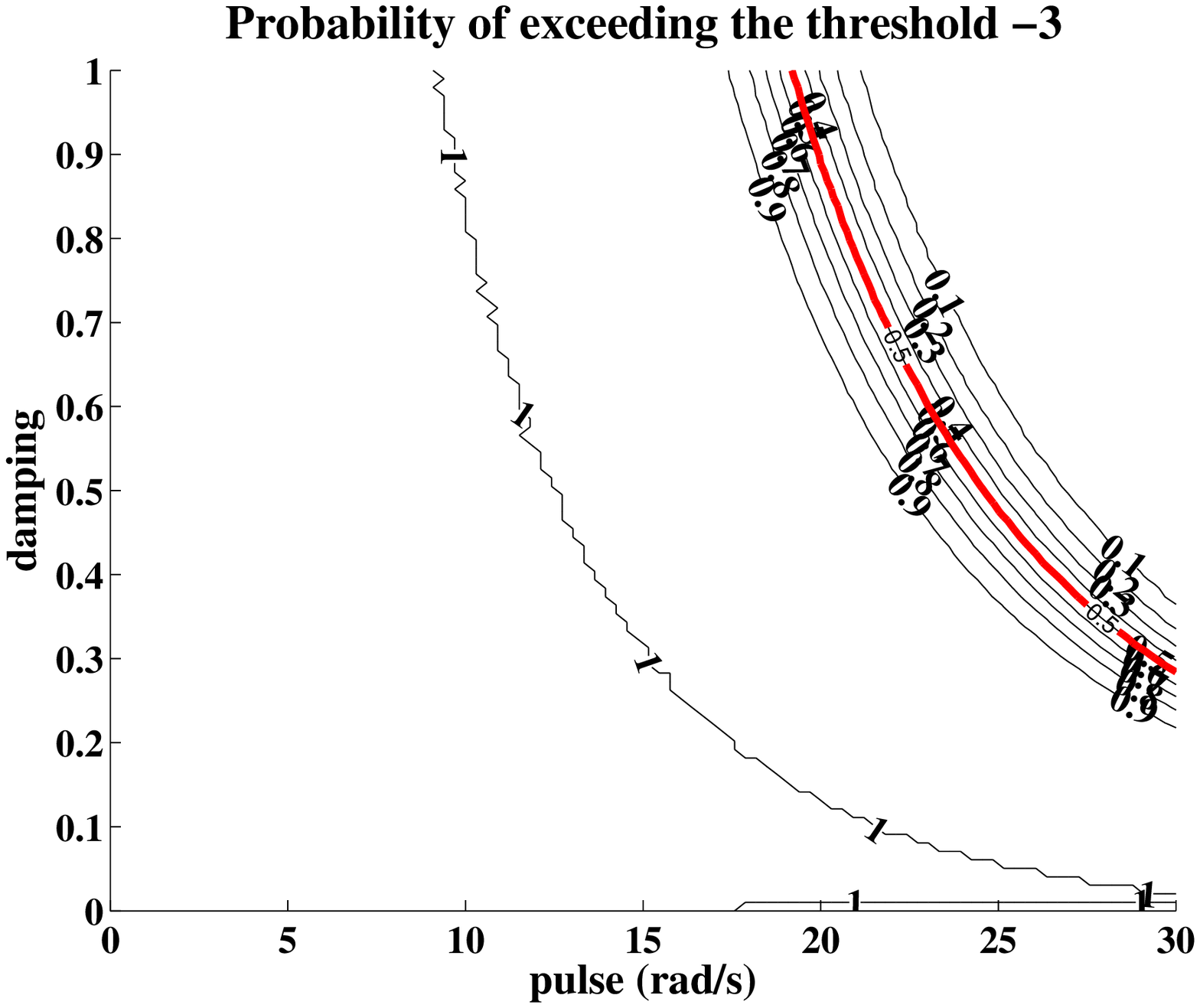}
\label{fig:function_probHF}}

\caption[Plot of the functions] {Our simulator $Z$ defined by
  \eqref{eq:function}.  Each column corresponds to a level of fidelity
  (from left to right, $\dt = 1$, $0.1$ and $\SI{0.01}{\second}$). First
  row corresponds to the mean function. Second row corresponds to the
  standard deviation function. Third row corresponds to the probability
  of exceeding the threshold.  The red bold line indicates the contour
  threshold.}
\label{fig:function}
\end{figure}

The pair $\left(\omega_0, \zeta\right)$ corresponds to the input vector
$x$ and the step time $\dt$ plays the role of the fidelity
parameter. The critical threshold $\zcrit$ is set to $-3$. %
The highest level of fidelity is set to $\thf = \SI{0.01}{\second}$. %
The computational cost is empirically linear with respect to the
fidelity level: $C\left(\dt\right) = a/\dt + b$, with $a = 0.0098$ and
$b = 0.02$ (coefficients chosen to have $C\left( \thf \right) = 1$). %
Figure~\ref{fig:function} represents the contour plots of the mean function,
the standard deviation and the probability function at three different
levels of fidelity.

We compute a reference value $\refer{\p}$ of $\p$ on a $100\times 100$
regular grid at the highest level of fidelity using $10^4$ simulations
of $Z$ (see Figure~\ref{fig:function_probHF}).  We suppose that the
input distribution $\Px$ is the uniform distribution on $\left[0;
  30\right]\times \left[0; 1\right]$ and let $\mu = \Px$.  The mean value $\refer{\globP} =
83.3\%$ of $\refer{\p}$ on the grid serves as a reference value for
$\globP$.

We use the Bayesian model of Section~\ref{sec:goal} with fixed
hyper-parameters, and ten fixed levels of fidelity: $\dt = 1$, 0.5,
0.33, 0.25, 0.2, 0.17, 0.1, 0.05, 0.02 and \SI{0.01}{\second}.  The
initial design (initial simulations before applying our \gls{msur} strategy)
consists of $180\times 60\times 20 \times 10 \times 5$ nested observations on
the five lowest fidelity levels (180 on the level $\dt =
\SI{1}{\second}$, 60 on the level $\dt = \SI{0.5}{\second}$, \dots). The
initial design is set using the algorithm of \citet{qian2009nested}.

In our experiments, we compare two strategies: five \gls{sl}
strategies and the \gls{msur} strategy \eqref{eq:MSUR}. %
For \gls{sl} strategies, all new points are sequentially selected on a
fixed level, using then \gls{sur} strategy \eqref{eq:surPrinc}. %
Each \gls{sl} strategy corresponds to one level of fidelity. %
Each time a new observation point must be selected, we choose the
point that achieves the best value of $J_n$ among 500 candidate points
per level drawn from~$\Px$. %
We allocate a simulation-time budget of~20 for each strategy. %
All experiments are repeated 12~times. %
Integrals are approximated by a sum on the $100\times 100$ regular
grid.

\begin{figure}
\centering

\psfrag{Cost}[tc][tc]{\footnotesize Cost}
\psfrag{E[P] - P}[bc][bc]
{\footnotesize $\sqrt{\aver{\left(\estim{\globP}_n
- \refer{\globP}\right)^2}}$}
\psfrag{||E[P(x)] - P(x)||}[bc][bc]
{\footnotesize $\sqrt{\aver{ \left\|\estim{p}_n
  - \refer{\p}\right\|^2_{\Ltwo} }}$}

\psfrag{Multi-fidelity}[cl][cl]{\footnotesize \gls{msur}}
\psfrag{Single level dt = 0.2}[cl][cl]
{\footnotesize \gls{sl} $\dt = \SI{0.2}{\second}$}
\psfrag{Single level dt = 0.1}[cl][cl]
{\footnotesize \gls{sl} $\dt = \SI{0.1}{\second}$}
\psfrag{Single level dt = 0.05}[cl][cl]
{\footnotesize \gls{sl} $\dt = \SI{0.05}{\second}$}
\psfrag{Single level dt = 0.02}[cl][cl]
{\footnotesize \gls{sl} $\dt = \SI{0.02}{\second}$}
\psfrag{Single level dt = 0.01}[cl][cl]
{\footnotesize \gls{sl} $\dt = \SI{0.01}{\second}$}

\psfrag{Error on the global probability}[bc][bc]{}
\psfrag{Error (L2) on the probability function}[bc][bc]{}

\subfloat[Error on $\globP$]
{\includegraphics[width = \widthFigure]{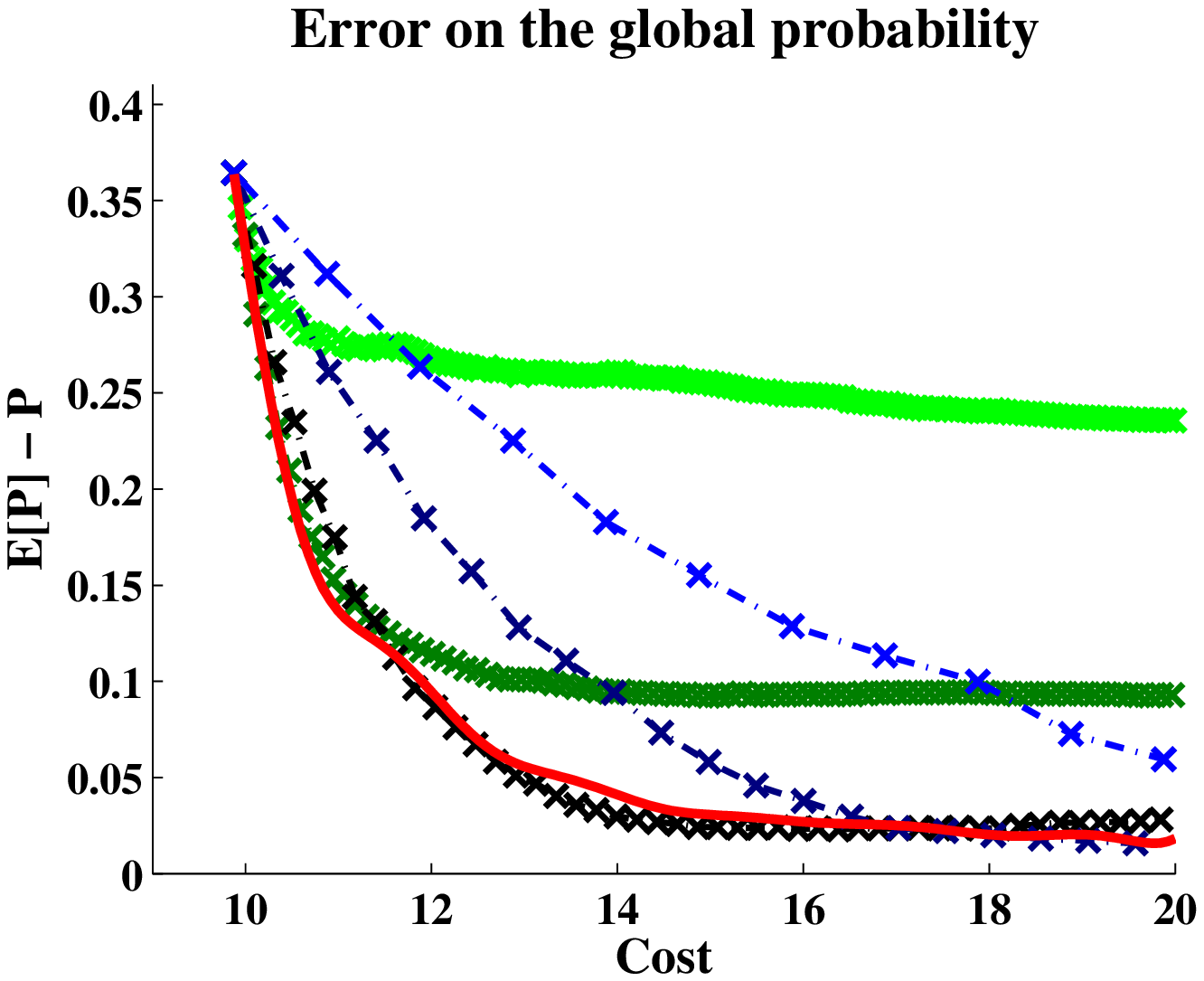}
\label{fig:results_error}}
\hspace{\intFigure}
\subfloat[Error on $\p$]
{\includegraphics[width = \widthFigure]{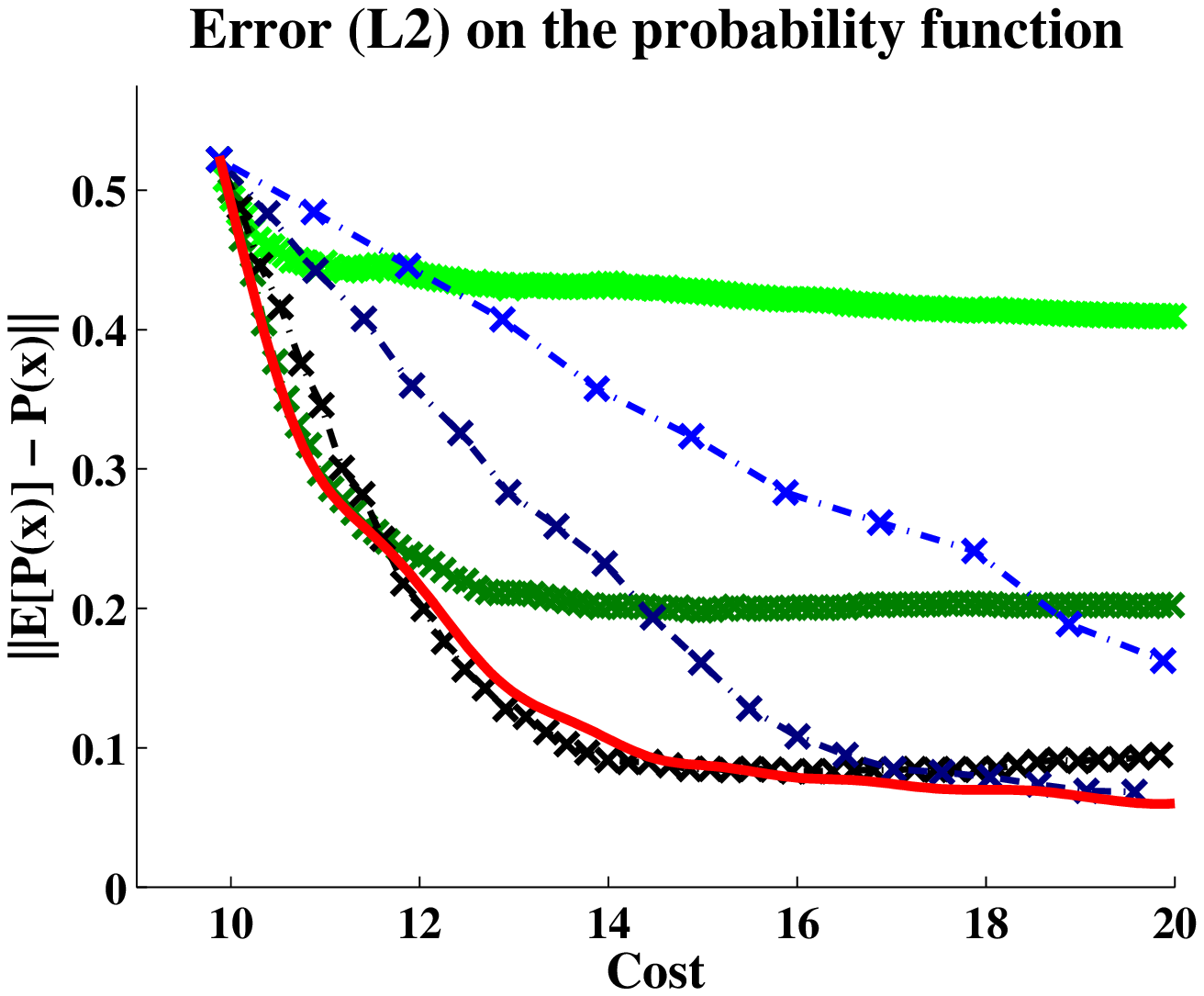}
\label{fig:results_criterion}}
\hspace{\intFigure}
\subfloat{\includegraphics[width = \widthFigure]{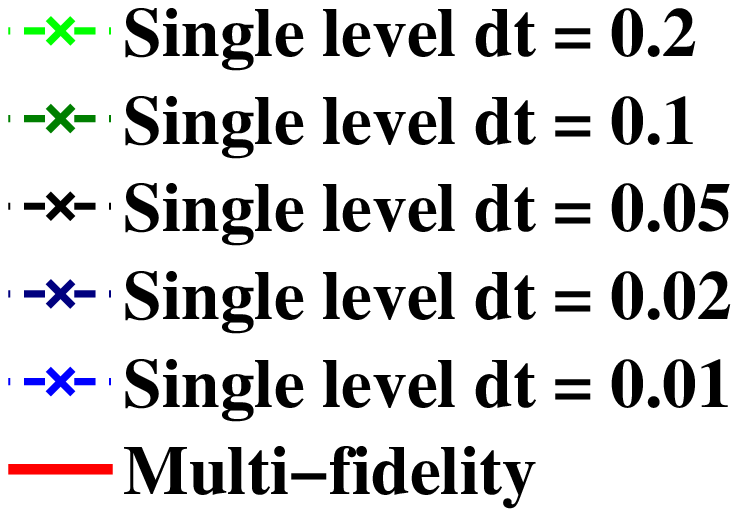}}

\caption{Mean square error of the quantities of interest
get with the \gls{msur} strategy and the \acrlong{sl} strategies.
The symbol $\aver{\cdot}$ means ``average on the 12 repetitions''.}
\label{fig:results}
\end{figure}

The strategies are compared based on the mean square error between
estimations and references. 
Results are presented on Figure~\ref{fig:results}.
Figure~\ref{fig:results_error} represents the mean square error of the
estimation $\estim{\globP}_n$ of $\globP$, and
Figure~\ref{fig:results_criterion} represents the mean square error of
the estimation $\estim{\p}_n$ of $\p$, as a function of the
computational cost.  Each curve corresponds to an average over $12$
experiments with the same method of sequential design.  The plain red
curve corresponds to the \gls{msur} strategy.  The dotted crossed
blue-green curves correspond to \gls{sl} strategies (one curve per
level). We can see that the best level is achieved at $\dt =
\SI{0.05}{\second}$.  The lower fidelity levels are too biased to
estimate the probabilities correctly, and the upper fidelity levels are
too expensive to make it possible to carry out enough observations in
order to estimate the probabilities accurately.  Moreover, we can see
that the \gls{msur} strategy is as good as the \gls{sl} strategy at $\dt =
\SI{0.05}{\second}$.  Consequently, with the \gls{msur} strategy, one does
not need to know which level yields the best trade-off between  accuracy
and computational cost.

\section{Conclusion} \label{sec:conclu}

This article makes two contributions. %
First, we suggest a new \gls{sur} criterion to estimate a probability
of exceeding a threshold in the case of a stochastic simulator. %
Second, we construct a sequential strategy called \gls{msur} as an
adaptation of the \gls{sur} strategies to deal with multi-fidelity
simulators. %
Our first results are promising, because the \gls{msur} strategy
succeeds to get the better of all single-level \gls{sur} strategies
without knowing which level provides the best compromise between speed
and accuracy.

\bibliographystyle{apalike}
\bibliography{ISIWCbibliography}

\end{document}